# Kinetics of the photochromic effect in oxygen-containing rare-earth hydrides

Dmitrii Moldarev,[1] Tuan T. Tran, Max Wolff, and Daniel Primetzhofer

*Department of Physics and Astronomy, Uppsala University, Box 516, 751 20 Uppsala, Sweden.*

The kinetics of the photochromic reaction of oxygen-containing rare-earth hydrides is commonly described by an exponential function assuming a single-step process. In this paper, we elaborate on the origin of the photochromic effect in oxygen-containing rare-earth metal hydrides, considering the kinetics of the reaction as a two-step process. We show that the fit to the experimental data is improved drastically when two processes that cause the photodarkening are assumed: a fast reaction rate-limited - for example, electronic or local – process and a slow, e.g. diffusion-limited process.

*Introduction* - Oxygen-containing rare-earth hydrides (REHO, RE =Y, Gd, Er, Dy, Sm, Sc, and Nd) stand out from other photochromic materials, as they exhibit a color-neutral photochromic effect under ambient conditions [1–4]. Being fairly transparent in the visible and near-IR parts of the spectrum (≈ 80%, see Figure 1), REHO films lose almost 50 % of their transmittance without a pronounced absorption band in the dark state - see Figure 1. This unique property makes REHO a promising candidate for applications in smart windows [5].

REHO films are typically produced using magnetron sputtering [6], although e⁻-beam evaporation can also be used as well [7]. It was demonstrated how variations in deposition conditions such as deposition pressure affect the microstructure [8], the chemical composition [9] and the optical properties of the films [10, 11]. Studies of YHO films of different thicknesses revealed a stronger photochromic response for thicker films, indicating that photochromism is a bulk effect [12, 13]. The films were proven to show the photochromic effect in different environments: UHV and O-free gas [14], under protective capping [15] and in tandem with another photochromic oxide layer [16, 17]. A stronger photochromic reaction, i.e. a larger contrast, was observed when the films are illuminated with light of a shorter wavelength [14].

The aforementioned findings can serve as a reasonable guide for device development. However, physical models proposed to explain photochromic properties in REHO are still being challenged. Moreover, the phase composition of REHO films is not fully understood. It has been proposed that REHO is single phase and belongs to the class of oxyhydrides, materials with mixed O⁻ and H⁻ anions [18] similar to oxyhydride powders prepared via topochemical reaction [19]. The appeal of this model is in its simplicity, and it allows using *ab initio* methods to study photochromic behavior [20–22]. Hans *et al.* debunked a single-phase concept demonstrating that films consist of O-rich and H-rich regions [23]. In the same work, the authors discussed the implications of the multiphase nature of REHO for the photochromic properties and proposed H migration as a mechanism of the photochromic reaction. Indeed, indications of H mobility and structural rearrangement under illumination have been shown in several studies [14, 24–26]. Earlier, the highest reported H⁻ conductivity has been demonstrated for a single-phase $LaH_{3−2x}O_x$ [27].

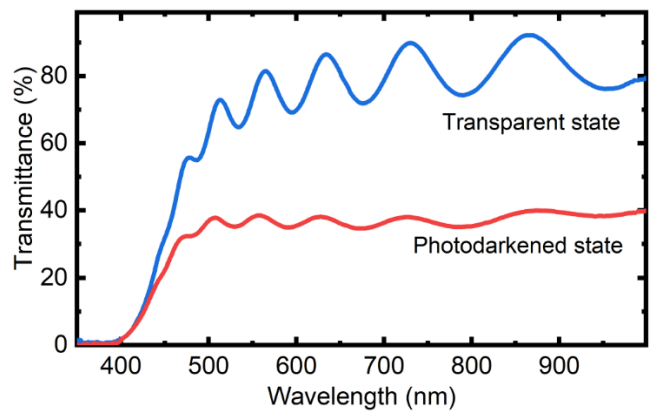

FIG. 1. Wavelength dependence of the optical transmittance of the YHO film in transparent (blue) and photodarkened (red) states. The color-neutral photochromism is apparent from the absence of a wavelength dependence in the entire visible spectrum. Intensity fringes are associated with reflection and interference.

In a recent paper, Z. Wu *et al.* investigated the time dependent evolution of defects in YHO under illumination employing positron annihilation lifetime spectroscopy and showed a correlation between the bleaching and the disappearance of metallic domains [28]. In this work, we elaborate on the mechanism of the photochromic reaction based on the nontrivial time dependence of the optical transmittance.

*Results and discussion* - Figure 2 shows the change in optical transmittance of the YHO film during 30 min of illumination with blue light and 30 min of bleaching. Fast decay is found during the first 2-3 minutes of illumination (≈ 30%) followed by slower darkening. A similar but inverse behavior is also characteristic for the relaxation process. Note that the bleaching for this sample is much slower than the photodarkening, which is commonly observed for YHO samples with strong photochromism. To study the kinetics of the photodarkening one can adopt an approach from analytic chemistry and consider the photochromic transition as a photochemical reaction, where reactants and products have different optical properties.

---

[1] Dmitry.Moldarev@physics.uu.se

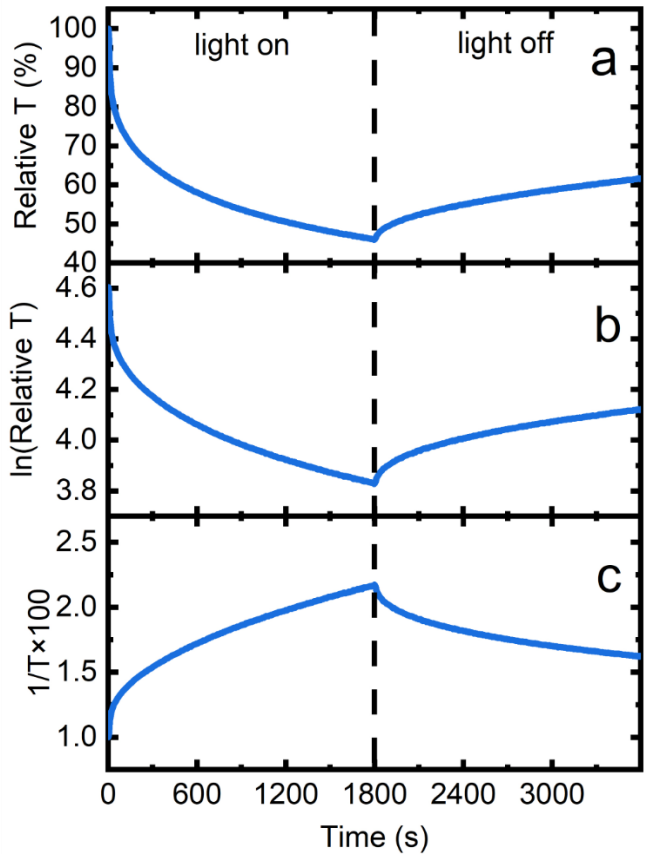

FIG. 2. Relative changes in the optical transmittance of YHO when light is switched on and off, plotted with (a) linear ordinate, (b) logarithmic ordinate, (c) reciprocal ordinate. The absence of linearity in any of the plots indicates a non-trivial time-dependence such as caused by multiple contributions to the effect with different time constants.

According to the Beer-Lambert law, optical attenuation is proportional to the concentration of attenuating species. The kinetics of changes in transmittance $T$ during the photochromic transition (see Figure 2) follows the kinetics of the reaction. Considering the simplest case, where one transparent reactant $A$ and one opaque product $B$ are involved in the reaction ($A \leftrightarrow B$), the kinetics can be described as a zero (Eq. 1), first (Eq. 2) or second order reaction (Eq. 3).

$$dC_A(t)/dt = -k_p; \qquad (1)$$
$$dC_A(t)/dt = -k_p C_A; \qquad (2)$$
$$dC_A(t)/dt = -k_p C_A^2. \qquad (3)$$

Here $k_p$ is the rate of the reaction, $C_A \propto T$ is the concentration of $A$. We may consider $A$ and $B$ as transparent and photodarkened states of YHO, respectively. In this case, $C_A$ represents a fraction of the film that has not undergone the photochromic transition.

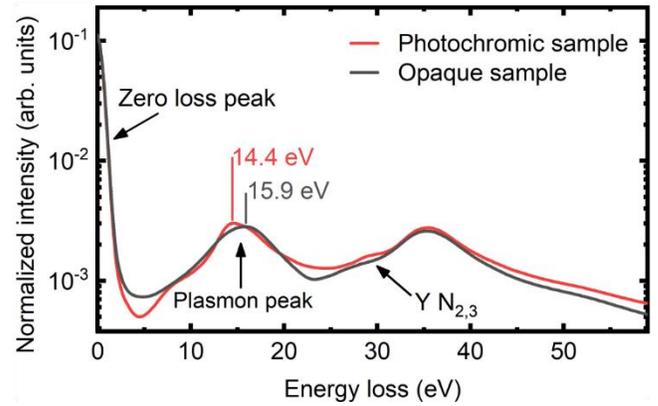

FIG. 3. EELS spectra for photochromic (red) and opaque (black) YHO films. Shift in the position of the plasmon peak (14.4 eV vs 15.9 eV) indicates a difference in H mobility.

Equations 1 and 2 yield linear and exponential time dependence, respectively, while Equation 3 provides a linear time dependence for $T^{-1}$. As seen in Figure 2, where the relative change in transmittance is plotted with linear, logarithmic, and reciprocal ordinates, none of these equations can successfully describe the photochromic transition.

REHO-based photochromic films have been found to be multiphase [23], therefore, a complex scheme of the transition, including interphase H diffusion, might be involved in the photochromic reaction.

To identify the presence of free and potentially mobile H, electron energy loss spectroscopy (EELS) has been employed for two YHO films: photochromic and opaque (inherently non-photochromic). The results are presented in Figure 3. The plasmon peak for the photochromic film is observed at 14.4 eV, while the opaque sample exhibits a plasmon peak at 15.9 eV. As discussed in Ref. [29], free hydrogen in its elemental form is detectable through an ionization edge at 13.6 eV, corresponding to transitions to continuum states of an isolated atom. However, when hydrogen is chemically bound to other elements, it donates electrons to the solid, eliminating characteristic ionization edges and typically shifting the plasmon peak upward by 1-2 eV due to increased valence-electron density. The 1.5 eV difference between the plasmon peaks for the two samples suggests a higher concentration of free or mobile hydrogen in the photochromic films comparing to the opaque sample. Free hydrogen is expected to be in the grain boundaries and micropores and can potentially facilitate the photochromic effect through light-induced hydrogen migration.

To take into account light-induced H migration, we can modify equation 2 (first-order reaction) adding a term that describes H diffusion into a volume, which undergoes a photochromic transition, from the rest of the film:

$$dC_A(t)/dt = -k_1 C_A + b e^{-k_2 t}, \qquad (4)$$

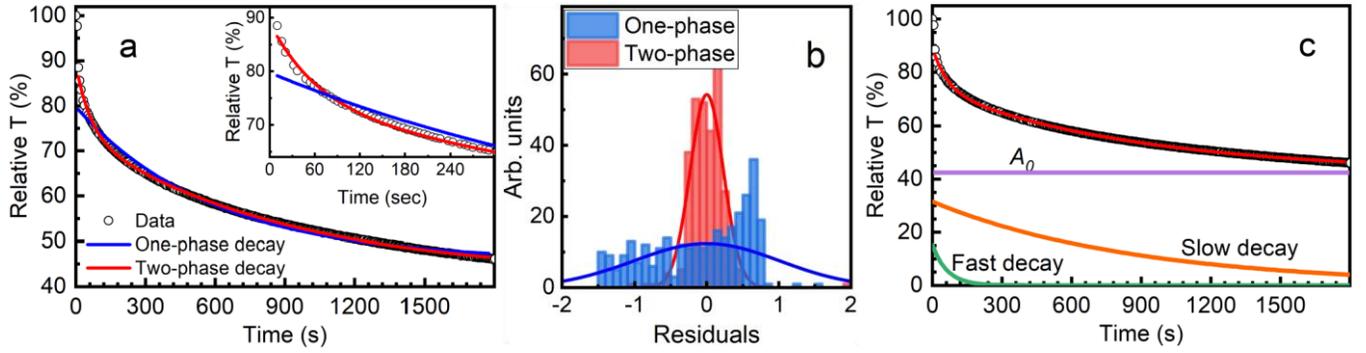

FIG. 4. (a) Relative changes in the optical transmittance of the YHO film under illumination and corresponding fit with one-phase exponential decay (blue) and two-phase exponential decay (red). The inset shows a close up of the first 5 minutes of illumination. (b) Density distribution of the residual between photodarkening data and the fit. Blue bars: fit using one-phase decay; red bars: fit using two-phase decay. Solid lines are Gaussian fits to the distributions. (c) Fit using two-phase exponential decay as a sum of three terms: constant $A_0$ (purple line), *fast* decay (green line) and *slow* decay (orange line).

where $k_2$ describes the rate of H transfer and $b$ is proportional to the amount of available H.

The solution of this differential equation yields a two-phase exponential decay. In addition, to our knowledge, YHO films have never been successfully photodarkened to a completely opaque state ($T$=0%). Therefore, we add a constant term $A_0$ that represents a volume fraction of the film that is not photoactive and cannot undergo the photochromic transition. The resulting equation is formulated as follows:

$$C_A(t) = A_0 + A_1 e^{-k_1 t} + A_2 e^{-k_2 t}. \qquad (5)$$

Here $A_1$ is a constant determined by initial conditions and $A_2 = b/(k_1 - k_2)$.

The result of the fit using a two-phase exponential decay function (Eq. 5) is presented in Figure 4a (red line) and compared to a single-phase function (blue line). The discrepancy between the two models is especially evident in the first minutes of illumination (see the inset in Figure 4a). The quality of the fit is drastically improved compared to a single-phase model as seen in Figure 4b, showing the density distribution of the residual using the two models. The two decays occur on a quite different time scale, as seen in Figure 4c, where each term of the two-phase exponential decay is presented as an individual function, implying that very different physics processes are involved. The *fast* decay takes place within the first 2-3 minutes ($k_1$=0.016 s$^{-1}$) followed by the *slow* decay governing the rest of the photodarkening ($k_2$=0.001 s$^{-1}$).

Time-resolved XRD measurement under illumination revealed a gradual lattice contraction that continues for at least 2-3 hours, i.e. as long as the measurement was going [26]. Therefore, the *slow* decay is probably attributed to hydrogen migration/diffusion within the film, locally and more distantly between domains, and its associated lattice contraction. The *fast* part, considering its time scale, can be associated with some local process, which could be driven directly by light absorption and does not require mass transport. Besides, fast changes in the first minutes can be related to loosely bound hydrogen leaving the film. In fact, H-loss during the first minutes of illumination has been demonstrated for different YHO films using two different vacuum systems [14, 24]. After the first minutes of photodarkening, no detectable H release was observed and $H_2$ partial pressure in the chamber started to decrease. Finding a way of tuning between the two processes (e.g. by modifying a microstructure) might allow one to tailor photochromic properties.

We acknowledge that there could be more factors that influence the photodarkening kinetics and that additional contributions could be taken into account. The bleaching reaction is a thermally accelerated process [30], therefore, it can also occur during illumination, competing with photodarkening. Additionally, because of the absorption of triggering light, the surface layer of the YHO film is exposed to light of higher intensity, which results in a nonuniform distribution of the newly formed absorbing species and deviation from Beer-Lambert law. Nevertheless, the proposed model with the two-phase decay function already provides excellent fitting quality.

*Materials and Methods* - Thin films of oxygen-containing yttrium hydride were deposited on 1 mm thick glass substrates using reactive magnetron sputtering in a Balzer Union sputtering system. An yttrium disc of 99.99 % purity was used as a sputter target. The base pressure of the system was below 5×10$^{-5}$ mbar prior to deposition. The pressure of a mixture of Ar and $H_2$ was kept at the level of 10$^{-2}$ mbar during the deposition process. The resulting hydride films were oxidized in air after venting the sputtering chamber.

The photochromic properties were characterized using a set-up allowing transmittance measurements under controlled environment [24]. The optical properties were measured using a stabilized tungsten-halogen light and a compact CCD spectrometer (ThorLabs CCS200). The photochromic reaction was initiated by a 405 nm LED with a nominal irradiance of 14.53 µW/mm$^2$. Transmittance was averaged over the wavelength range 500-900 nm and studied as a function of time.

*Conclusions* - The time dependence of the optical transmittance of YHO under illumination shows unambiguously two main processes, with very different time constants, contributing to the photochromic effect. The *fast* process is responsible for the rapid photodarkening that occurs in the first minutes of illumination and is probably associated with changes in the local structure. The gradual *slow* decay can be attributed to hydrogen migration/diffusion in the film. While the exact origin of the two processes can be further elucidated, the proposed model describes the experimental data well and allows a straightforward comparison of the performance of photochromic films. The latter is especially important when developing and optimizing photochromic devices for energy efficient applications.

ACKNOWLEDGMENTS

The authors acknowledge financial support of the project by the Olle Engkvist Foundation under grant number 207-0423. Infrastructure support by VR-RFI [grant #2019-00191]; supporting accelerator operation is gratefully acknowledged.